\documentclass[aps,prl,preprint,tightenlines,superscriptaddress,showpacs,byrevtex]{revtex4}
\usepackage{graphicx} 
\usepackage{dcolumn}  
\graphicspath{{ps}}

\begin{document}

\preprint{\vbox{ \hbox{   }
                 \hbox{BELLE-CONF-0655}
}}


\title{ \quad\\[0.5cm]
The first observation of $\tau^{\pm} \to \phi K^{\pm} \nu$ decay
}

\affiliation{Budker Institute of Nuclear Physics, Novosibirsk}
\affiliation{Chiba University, Chiba}
\affiliation{Chonnam National University, Kwangju}
\affiliation{University of Cincinnati, Cincinnati, Ohio 45221}
\affiliation{University of Frankfurt, Frankfurt}
\affiliation{The Graduate University for Advanced Studies, Hayama} 
\affiliation{Gyeongsang National University, Chinju}
\affiliation{University of Hawaii, Honolulu, Hawaii 96822}
\affiliation{High Energy Accelerator Research Organization (KEK), Tsukuba}
\affiliation{Hiroshima Institute of Technology, Hiroshima}
\affiliation{University of Illinois at Urbana-Champaign, Urbana, Illinois 61801}
\affiliation{Institute of High Energy Physics, Chinese Academy of Sciences, Beijing}
\affiliation{Institute of High Energy Physics, Vienna}
\affiliation{Institute of High Energy Physics, Protvino}
\affiliation{Institute for Theoretical and Experimental Physics, Moscow}
\affiliation{J. Stefan Institute, Ljubljana}
\affiliation{Kanagawa University, Yokohama}
\affiliation{Korea University, Seoul}
\affiliation{Kyoto University, Kyoto}
\affiliation{Kyungpook National University, Taegu}
\affiliation{Swiss Federal Institute of Technology of Lausanne, EPFL, Lausanne}
\affiliation{University of Ljubljana, Ljubljana}
\affiliation{University of Maribor, Maribor}
\affiliation{University of Melbourne, Victoria}
\affiliation{Nagoya University, Nagoya}
\affiliation{Nara Women's University, Nara}
\affiliation{National Central University, Chung-li}
\affiliation{National United University, Miao Li}
\affiliation{Department of Physics, National Taiwan University, Taipei}
\affiliation{H. Niewodniczanski Institute of Nuclear Physics, Krakow}
\affiliation{Nippon Dental University, Niigata}
\affiliation{Niigata University, Niigata}
\affiliation{University of Nova Gorica, Nova Gorica}
\affiliation{Osaka City University, Osaka}
\affiliation{Osaka University, Osaka}
\affiliation{Panjab University, Chandigarh}
\affiliation{Peking University, Beijing}
\affiliation{University of Pittsburgh, Pittsburgh, Pennsylvania 15260}
\affiliation{Princeton University, Princeton, New Jersey 08544}
\affiliation{RIKEN BNL Research Center, Upton, New York 11973}
\affiliation{Saga University, Saga}
\affiliation{University of Science and Technology of China, Hefei}
\affiliation{Seoul National University, Seoul}
\affiliation{Shinshu University, Nagano}
\affiliation{Sungkyunkwan University, Suwon}
\affiliation{University of Sydney, Sydney NSW}
\affiliation{Tata Institute of Fundamental Research, Bombay}
\affiliation{Toho University, Funabashi}
\affiliation{Tohoku Gakuin University, Tagajo}
\affiliation{Tohoku University, Sendai}
\affiliation{Department of Physics, University of Tokyo, Tokyo}
\affiliation{Tokyo Institute of Technology, Tokyo}
\affiliation{Tokyo Metropolitan University, Tokyo}
\affiliation{Tokyo University of Agriculture and Technology, Tokyo}
\affiliation{Toyama National College of Maritime Technology, Toyama}
\affiliation{University of Tsukuba, Tsukuba}
\affiliation{Virginia Polytechnic Institute and State University, Blacksburg, Virginia 24061}
\affiliation{Yonsei University, Seoul}
  \author{K.~Abe}\affiliation{High Energy Accelerator Research Organization (KEK), Tsukuba} 
  \author{K.~Abe}\affiliation{Tohoku Gakuin University, Tagajo} 
  \author{I.~Adachi}\affiliation{High Energy Accelerator Research Organization (KEK), Tsukuba} 
  \author{H.~Aihara}\affiliation{Department of Physics, University of Tokyo, Tokyo} 
  \author{D.~Anipko}\affiliation{Budker Institute of Nuclear Physics, Novosibirsk} 
  \author{K.~Aoki}\affiliation{Nagoya University, Nagoya} 
  \author{T.~Arakawa}\affiliation{Niigata University, Niigata} 
  \author{K.~Arinstein}\affiliation{Budker Institute of Nuclear Physics, Novosibirsk} 
  \author{Y.~Asano}\affiliation{University of Tsukuba, Tsukuba} 
  \author{T.~Aso}\affiliation{Toyama National College of Maritime Technology, Toyama} 
% ...
\collaboration{The Belle Collaboration}


\begin{abstract}
We present the first measurement of tau-decays to hadronic final 
states with a $\phi$-meson.
This is based on 401.4 fb$^{-1}$ of data accumulated at the Belle experiment. 
The branching ratio obtained is
$B(\tau^{\pm}\to\phi K^{\pm}\nu) = (4.06\pm 0.25\pm 0.26)\times 10^{-5}$.
\end{abstract}



\maketitle
\tighten

\section{Introduction}

While hadronic $\tau^{\pm}$ decays
with a $\phi$ meson in the final state
are valuable to investigate QCD at a low mass scale, 
they have never been observed due to their small branching fractions. 
The decay $\tau^{\pm} \to \phi K^{\pm} \nu$ is Cabibbo suppressed and 
further restricted by its small phase space, while the decay 
$\tau^{\pm} \to \phi \pi^{\pm} \nu$ is suppressed 
by the OZI rule although it is Cabibbo allowed. 
Taking into accounts differences in Cabibbo mixing effects and phase space
relative to the Cabibbo allowed transition $\tau^{\pm} \to K^{0*} K^{\pm} \nu$, 
the branching fraction for $\tau^{\pm} \to \phi K^{\pm} \nu$
is estimated to be 
$B(\tau^{\pm} \to \phi K^{\pm} \nu)= 2\times 10^{-5}$~\cite{CLEO}. 
On the other hand, the vector dominance model predicts 
$B(\tau^{\pm} \to \phi \pi^{\pm} \nu) = (1.20 \pm 0.48)\times10^{-5}$~\cite{Castro}. 

CLEO searched for these decays using 3.1 fb$^{-1}$ of data taken on $\Upsilon(4S)$ 
resonance. They set upper limits of
$B(\tau^{\pm} \to \phi K^{\pm} \nu)<(5.4-6.7)\times 10^{-5}$ and  
$B(\tau^{\pm} \to \phi \pi^{\pm} \nu)<(1.2-2.0)\times 10^{-4}$ at 
the 90\% confidence level~\cite{CLEO}.
Here we report the first measurement of $\tau^{\pm} \to \phi K^{\pm} \nu$ and 
$\tau^{\pm} \to \phi K^{\pm}\pi^0 \nu$ decays.
These results are based on a data sample of 401.4 fb$^{-1}$ 
corresponding to $3.6 \times 10^8$ $\tau^+ \tau^-$ pairs
collected near the $\Upsilon(4S)$ resonance
with the Belle detector at the KEKB asymmetric-energy $e^+e^-$ 
(3.5 on 8~GeV) collider~\cite{KEKB}. 
The Belle detector is a large-solid-angle magnetic
spectrometer that
consists of a silicon vertex detector (SVD),
a 50-layer central drift chamber (CDC), an array of
aerogel threshold \v{C}erenkov counters (ACC), 
a barrel-like arrangement of time-of-flight
scintillation counters (TOF), and an electromagnetic calorimeter
comprised of CsI(Tl) crystals (ECL) located inside 
a super-conducting solenoid coil that provides a 1.5~T
magnetic field.  An iron flux-return located outside of
the coil is instrumented to detect $K_L^0$ mesons and to identify
muons (KLM).  The detector is described in detail elsewhere~\cite{Belle}.
Two inner detector configurations were used. A 2.0 cm beampipe
and a 3-layer silicon vertex detector was used for the first sample
of 157.7 fb$^{-1}$, while a 1.5 cm beampipe, a 4-layer
silicon detector and a small-cell inner drift chamber were used to record  
the remaining 243.7 fb$^{-1}$~\cite{Ushiroda}.  

While the decay $\tau^{\pm}\to\phi\pi^{\pm}\nu$ is also interesting, 
it is treated here as a background process including the kinematically 
allowed but phase-space suppressed decays of 
$\tau^{\pm}\to\phi (n \pi) \nu$ ($2 \leq n\leq 5)$.

\section{Event Selection}

We look for $\tau^{\pm}\to\phi K^{\pm}\nu$ candidates in
$e^+e^-\to \tau^+\tau^-$ reaction in the following modes.

\hspace*{30 mm}$\tau^{\pm}_{\rm signal} \to \phi + K^{\pm} + (\rm{missing})$ \\
\hspace*{46 mm}$\hookrightarrow K^+ K^-$ \\
\hspace*{40 mm}$\tau_{\rm tag}^{\mp} \to (\mu/ e)^{\mp} + n (\leq 1) \gamma + 
(\rm{missing})$ \\

\vspace*{-3 mm}
The detection of $\phi$-mesons relies on the $\phi \to K^+ K^-$ decay 
($B=(49.1\pm 0.6)\%$); the final evaluation of the signal yield
is carried out using the $K^+ K^-$ invariant mass distribution. 

Selection criteria described below are determined, based on examinations of 
Monte-Carlo (MC) events.
The background samples consist of
$\tau\tau$ (1570.0 fb$^{-1}$, which does not include any decay-mode 
with a $\phi$ meson) and other backgrounds such as 
$q\overline{q}$ continuum, $B^0{\overline{B}}^0$, $B^+ B^-$
and two-photon processes. 
For signal, we generate samples with $1\times 10^6$
$\tau^{\pm} \to \phi K^{\pm}\nu$, 
$\phi K^{\pm}\pi^0\nu$, $\phi\pi^{\pm}\nu$ and $\phi\pi^{\pm}\pi^0\nu$.

Transverse momentum ($p_t$) for a charged track is required to be larger than 
0.06 GeV/$c$ in the barrel region ($-0.6235<\cos \theta<0.8332$, where 
$\theta$ is the polar angle 
opposite to the incident $e^+$ beam direction 
in the laboratory frame)
and 0.1 GeV/$c$ in the endcap region ($-0.8660<\cos \theta<-0.6235$, and 
$0.8332<\cos \theta<0.9563$).
The energies of photon candidates are
required to be larger than 0.1 GeV in both regions. 

To select $\tau$-pair samples,
we require four charged tracks with zero net charge
and a total energy of charged tracks and photons
in the center-of-mass (CM) frame less than 11 GeV.
Furthermore, the missing momentum in the laboratory frame is required to be 
greater than 0.1 GeV/$c$ and its direction to be within the detector acceptance, 
where the missing momentum is defined as the deficit of sum of the observed 
momentum vectors from that of the initial $e^+e^-$ system. 
We also require $\cos \theta^{\rm CM}_{\rm thrust - miss}<-0.6$ to reduce 
$q\bar{q}$ backgrounds, where $\theta^{\rm CM}_{\rm thrust - miss}$ is an opening 
angle between the thrust axis and the missing momentum in CM frame. 
The event is subdivided into 3-prong and 1-prong
hemispheres according to the thrust axis 
in the CM frame. These are referred to
as the signal and tag side, respectively.

In order to remove the dominant $q\bar{q}$ background,
we require that the lepton probability $P_{\mu/ e}$ be greater than 0.1
and that
the invariant mass of the particles on
the tag side be less than 1.8 GeV/$c^2$ 
($\simeq m_{\tau}$). 
Similarly, we require that both kaon daughters of the $\phi$
candidate have kaon probabilities $P_K > 0.8$. The
effective mass of the signal side must be less
than 1.8 GeV/$c^2$. Here
$P_{\ell}$ is the likelihood that a charged particle
is of type ${\ell}$ ($\ell = \mu$ or $e$ or $K$ or $\pi$), defined as 
$P_{\ell} = L_{\ell}/(L_{\ell}+L_{x})$, 
where $L_{\ell}$ and $L_x$ are the likelihoods of the particle for $\ell$ 
and other species hypotheses, respectively, determined from responses of 
the relevant detectors. 
We allow at most one photon on the tag side
to take account of initial state radiation,
while requiring no extra photons on the signal side. 

Candidates $\phi$ mesons are reconstructed using 
oppositely charged kaons
within the barrel and forward endcap region.  
To suppress combinatorial backgrounds from other $\tau$ decays
and $q\bar{q}$ processes,
we require that the $\phi$ momentum be greater than 1.5 GeV/$c$ in the CM frame.
After these requirements the remaining contributions
from $B^0\bar{B}^0$, $B^+ B^-$, Bhabha, $\mu$ pair and 
two photon backgrounds are negligible.

To separate $\phi K^{\pm}\nu$ from $\phi\pi^{\pm}\nu$, the remaining charged 
track is required to satisfy the same kaon identification
criteria as the $\phi$ daughters.
The $\tau\tau$ and $q\bar{q}$ contributions are reduced by requiring 
the opening angle ($\theta^{\rm CM}_{\phi K}$) between $\phi$ and $K^{\pm}$
in the CM frame 
to satisfy $\cos\theta^{\rm CM}_{\phi K}>0.92$ and the 
momentum of the $\phi K^{\pm}$ system 
in the CM frame must be greater than 3.5 GeV/$c$.
For $\phi \pi^{\pm}\nu$, we require the charged track to be identified 
as a pion, $P_{\pi} > 0.8$,
and the opening angle between $\phi$ and $\pi^{\pm}$ in the
CM frame to satisfy $\cos\theta^{\rm CM}_{\phi \pi}< 0.98$. 

\begin{figure}[htb]
\centerline{\resizebox{10cm}{!}{\includegraphics{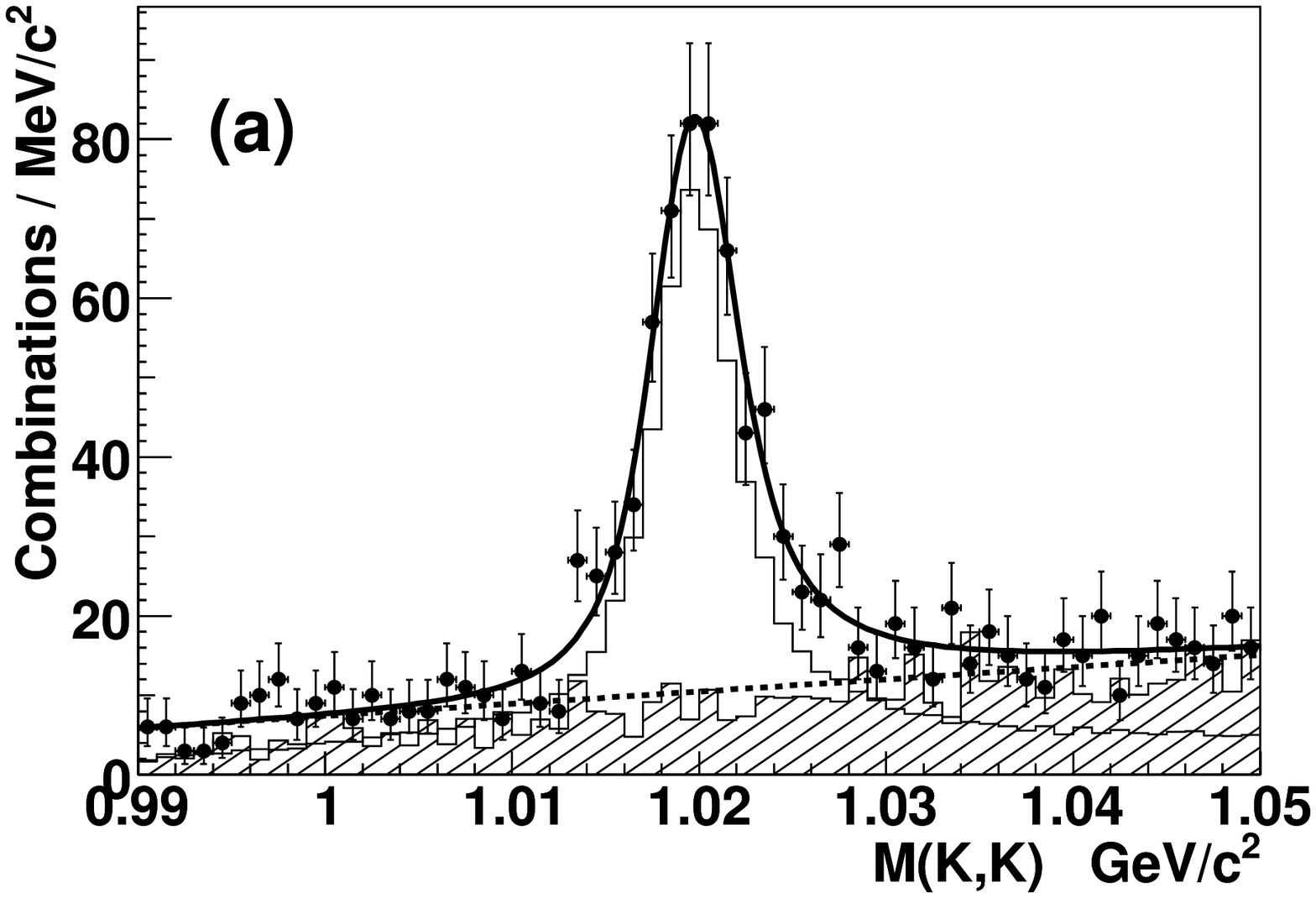}}}
\centerline{\resizebox{10cm}{!}{\includegraphics{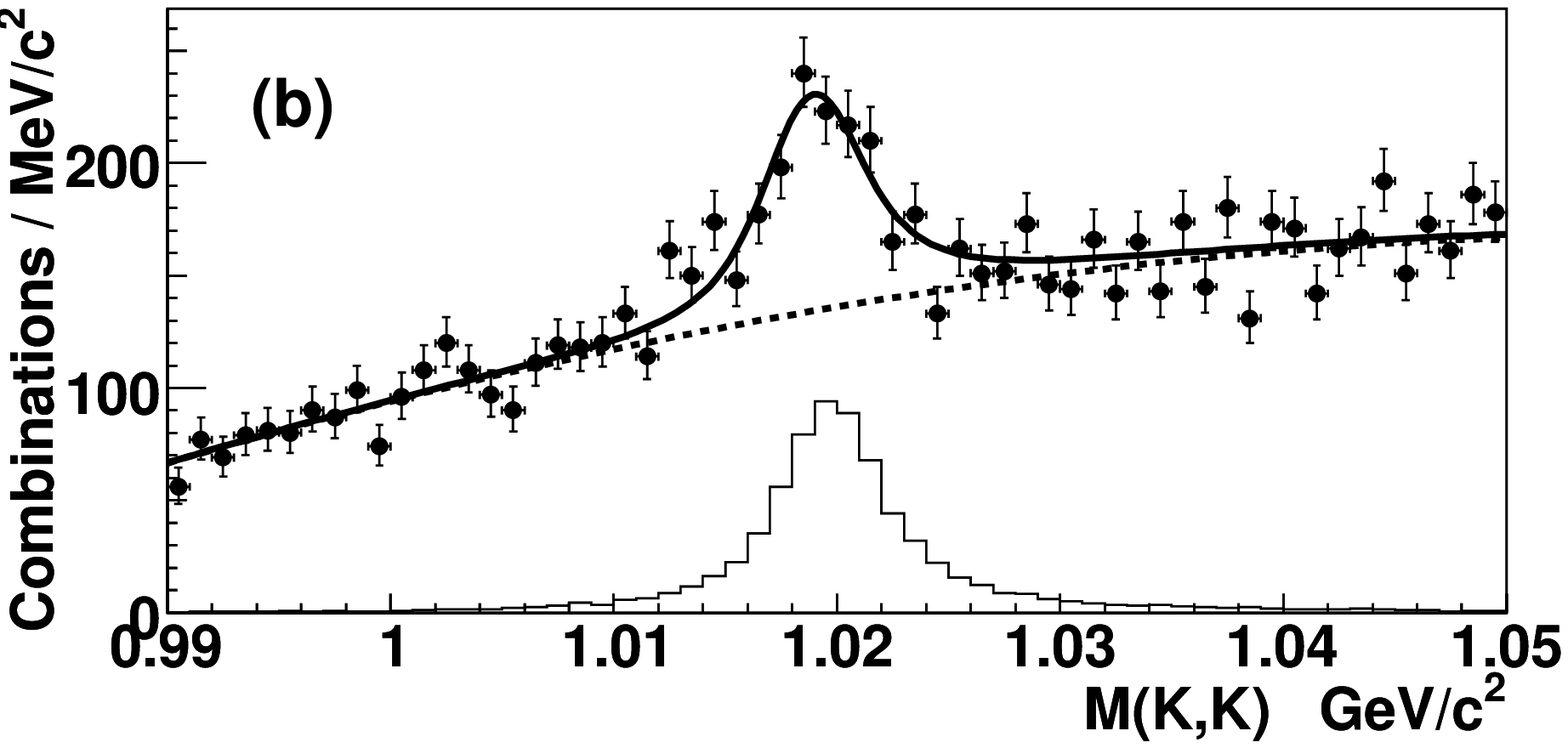}}}
\caption{$K^+ K^-$ invariant mass distributions for
(a) $\tau^\pm \to \phi K^\pm \nu$ and (b) $\tau^\pm \to \phi \pi^\pm \nu$.
Points with error bars indicate the data.
The hatched histograms show the 
expectations from $\tau^+\tau^-$ and $q\bar{q}$ background MCs.
The open histogram is the signal MC with 
B($\tau^\pm \to \phi K^\pm \nu$)$=4 \times 10^{-5}$ in (a) and
B($\tau^\pm \to \phi \pi^\pm \nu$)$=6 \times 10^{-5}$ in (b).
The curves show the best fit results. See the text for details.}
\label{fig:fdata}
\end{figure}

Fig.\ref{fig:fdata}(a)
shows the $K^+ K^-$ invariant mass
distribution after all $\tau^{\pm} \to \phi K^{\pm}\nu$
selection requirements.
While there are two possible $K^+ K^-$ combinatorial
contributions from 
$K^{\pm} K^{\mp} K^{\pm}$ on the signal side 
in forming a $\phi$ meson; all combinations are included.
Non-resonant backgrounds are mostly attributed to 
$\tau^{\pm} \to K^+ K^- \pi^{\pm} \nu$ events 
with $B = (1.55\pm 0.07)\times 10^{-3}$. Very
small contributions are expected
from $q\bar{q}$ processes. 

\section{Signal and background evaluation}

The detection efficiencies $\epsilon$ for $\tau^{\pm} \to \phi K^{\pm}\nu$, 
$\phi K^{\pm}\pi^0\nu$ and $\phi\pi^{\pm}\nu$ are evaluated, as listed in Table~\ref{tbl:eff}, 
from MC using KKMC~\cite{KKMC}, where the $V-A$ interaction is 
assigned on the vertices and the final hadrons decay according
to non-resonant phase space.
The signal $\tau^{\pm} \to \phi K^{\pm}\nu$ detection efficiency is 
$\epsilon_{\phi K\nu} = (1.82\pm 0.01)\%$, including the branching fraction of 
$\phi \to K^+K^-$. 

\begin{table}[h]
\begin{center}
\caption{Detection efficiencies $\epsilon$ and cross-feed rates (\%).} 
\label{tbl:eff}
\begin{tabular}{c|ccc}\hline
  & \multicolumn{3}{c}{Decay modes} \\ 
Candidates & $\phi K\nu$ & $\phi\pi\nu$ & $\phi K\pi^0\nu$ \\ \hline
$\tau \to \phi K\nu$ & 1.823$\pm$0.009 & 0.049$\pm$ 0.002 & 
0.327$\pm$0.006 \\
$\tau \to \phi \pi\nu$ & 0.110$\pm$0.002 & 1.660$\pm$ 0.014 & 
0.009$\pm$0.001 \\ \hline
\end{tabular}
\end{center}
\end{table}

Signal yields are evaluated by a fit to a p-wave Breit-Wigner (BW) distribution,
convoluted with a Gaussian function 
(of width $\sigma$) to account for the detector resolution.
The $\phi$ width is fixed to be $\Gamma_{\phi} = 4.26$ MeV 
(PDG value~\cite{PDG}), but $\sigma$ is allowed to float.
From fits to signal MC, the $\phi$ mass detector
resolution is found to be,
$\sigma=1.07\pm0.03$ MeV/$c^2$, which is consistent with
the fit result to the data of $1.2\pm0.3$ MeV/$c^2$.

The $K^+ K^-$ invariant mass distributions
for $\tau^{\pm} \to \phi K^{\pm}\nu$ and $\phi\pi^{\pm}\nu$ 
candidates are fitted with a p-wave BW distribution plus 
a linear and a second order polynomial background function, respectively, 
as seen in Figs.~\ref{fig:fdata}(a) and (b). 
The obtained signal yields are 
$N_{\tau \to \phi K\nu} = 573 \pm 32$ and 
$N_{\tau \to \phi \pi\nu} = 753 \pm 84$.

MC studies show that only $\tau^{\pm} \to \phi\pi^{\pm}\nu$,
$\tau^{\pm} \to \phi K^{\pm}\pi^0 \nu$ and $q\bar{q}$ samples yield 
significant contributions
peaking at the $\phi$ mass. The contribution of other
backgrounds is less than 0.01\% and can be neglected.
The number of $\tau^{\pm} \to \phi\pi^{\pm}\nu$ events in the data
is already evaluated. 
Other contributions are estimated below.

\begin{figure}[h]
\centerline{\resizebox{10cm}{!}{\includegraphics{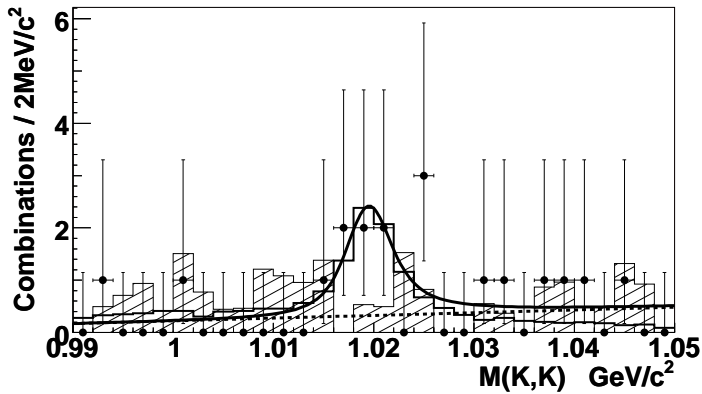}}}
\caption{$K^+ K^-$ invariant mass
distributions for $\tau^\pm \to \phi K^\pm \pi^0\nu$.
Points with error bars indicate the data.
Histograms show the MC
expectations of $\tau$-pairs (hatched) and signal (white) with 
a branching ratio of $3\times 10^{-6}$.
}
\label{phiKpi0}
\end{figure}

In order to evaluate the branching fraction and background
contribution from $\tau^{\pm} \to \phi K^{\pm}\pi^0\nu$,
we select $\pi^0 ( \to \gamma\gamma)$ candidates and combine them with
$\phi K^\pm \nu$ combinations that satisfy the requirements listed above.
The signal yield is estimated by fitting its 
$K^+ K^-$ invariant mass distribution with 
a p-wave BW distribution plus a linear background function,
as shown in Fig.\ref{phiKpi0}.
The resulting yield is $8.2\pm 3.8 \phi K \pi^0 \nu$ events in
data.
With the detection efficiency of $\epsilon_{\phi K\pi^0\nu} = 
(0.395\pm 0.007)\%$ evaluated by MC and the produced number of $\tau\tau$'s, 
$N_{\tau\tau} = 401.4$ (fb$^{-1}$) $\times 0.892$ (nb) $= 3.58\times 10^{8}$, 
we obtain a branching fraction of 
\begin{eqnarray}
B(\tau^{\pm} \to \phi K^{\pm}\pi^0\nu) = (2.9\pm1.3\pm0.2)\times 10^{-6}
\end{eqnarray} 
with a systematic uncertainty of 6.9\%,
which is described later.
Using this we estimate the contamination of $\tau^{\pm} \to \phi K^{\pm}\pi^0\nu$ events in
the $\tau^{\pm} \to \phi K^{\pm}\nu$ signal as
$N_{\tau^{\pm} \to \phi K^{\pm}\pi^0\nu}^{\rm cont} = (6.8\pm 3.1)$ events,
given the
cross-feed rate of $\tau^{\pm} \to \phi K^{\pm}\pi^0\nu$ to the
$\tau^{\pm} \to \phi K^{\pm}\nu$ signal to be $(0.327\pm 0.006)\%$, as listed in Table~\ref{tbl:eff}. 

From the MC (731.1 fb$^{-1}$) study, we find the contamination of $q\bar{q}$ is
$N_{q\bar{q}} = 12.1\pm 4.5$.
To take into account the uncertainty in $\phi$ production
in $q\bar{q}$ MC,
we compare MC results with enriched $q\bar{q}$ data by demanding 
that the effective 
mass of the tag side be larger than 1.8 GeV/$c^2$. 
With this selection, the background is $q\bar{q}$ dominated and
the other backgrounds are negligible. 
The yield in data is $262\pm 21$ events and 
$117.4\pm 9.9$ events for $q\bar{q}$. We then scale the
above estimate by the factor, $f = 2.23 \pm 0.47$, and obtain
the $q\bar{q}$ contamination,
$N_{q\bar{q}}^{\rm cont} = 14.8\pm 3.5$ events.

\section{Results}

The peaking backgrounds described above, 
$\tau^{\pm} \to \phi K^{\pm}\pi^0\nu$ and $q\bar{q}$, are subtracted 
from the signal yield,
$N_{\tau^{\pm} \to \phi K^{\pm}\nu} = (573\pm32)-(6.8\pm3.1)-(14.8\pm3.5)
= 551.4\pm32.3 ~~{\rm events}$.
To take into account cross-feed between 
$\tau^{\pm} \to \phi K^{\pm}\nu$ and $\tau^\pm \to \phi\pi^{\pm}\nu$
due to particle misidentification ($K\leftrightarrow \pi$), 
we solve the following simultaneous equations, 
\begin{eqnarray}
N_{\phi K\nu} &=& 2 N_{\tau\tau} \left( 
\epsilon_{\phi K\nu} \times B_{\phi K\nu} + 
\epsilon_{\phi \pi\nu}^{\phi K\nu} \times B_{\phi \pi\nu} 
\right), \\
N_{\phi \pi\nu} &=& 2 N_{\tau\tau} \left( 
\epsilon_{\phi K\nu}^{\phi\pi\nu} \times B_{\phi K\nu} + 
\epsilon_{\phi \pi\nu} \times B_{\phi \pi\nu}
\right), 
\end{eqnarray}
where $B_{\phi K\nu}$ and $B_{\phi \pi\nu}$ is the branching fraction for 
$\tau^{\pm} \to \phi K^{\pm} \nu$ and $\tau^{\pm} \to \phi \pi^{\pm} \nu$, 
respectively. 
$\epsilon$'s are the detection efficiencies listed in Table~\ref{tbl:eff}.
$\epsilon_{\phi K\nu}^{\phi\pi\nu}$
is the efficiency for reconstructing
$\tau^{\pm} \to \phi K^{\pm}\nu$ as 
$\tau^{\pm} \to \phi\pi^{\pm}\nu$ while
$\epsilon_{\phi \pi\nu}^{\phi K\nu}$
is the efficiency for reconstructing
$\tau^{\pm} \to \phi \pi^{\pm}\nu$ as $\tau^{\pm} \to \phi K^{\pm}\nu$. 
The resulting branching fraction for $\tau^{\pm} \to \phi K^{\pm}\nu$ is 
\begin{eqnarray}
B_{\phi K^{\pm}\nu} = (4.06 \pm 0.25) \times 10^{-5},
\end{eqnarray}
where the uncertainty is calculated with only statistical ones of $N_{\phi K\nu}$ and 
$N_{\phi\pi\nu}$. 
The uncertainty in the detection efficiencies, 
$\epsilon$'s, will be taken into account in the systematic error.
$B_{\phi \pi\nu}$ is obtained by the same way as 
\begin{eqnarray}
B_{\phi \pi\nu} = (6.07 \pm 0.71) \times 10^{-5},
\end{eqnarray}
but this is not the final branching fraction for the decay
since the unknown contamination of $\tau\to\phi (n\pi) \nu$ $(n\leq 5)$ 
decays still must be subtracted.

Systematic uncertainties are estimated as follows: Evaluation uncertainties of 
the integrated luminosity, $\tau\tau$ cross-section and trigger efficiency are 
1.4\%, 1.3\% and 1.1\%, respectively. 
Track finding efficiency has an uncertainty of 4.0\%. 
Uncertainties in lepton and kaon identification efficiencies and fake rate
are evaluated, respectively, to be 3.2\% and 3.1\% by averaging the
estimated uncertainties depending on momentum and polar angle of each charged track. 
To evaluate the systematic uncertainty of fixing $\Gamma_{\phi}$ in the BW fit, 
we calculate the change in the signal yield when
$\Gamma_{\phi}$ is varied
by $\pm 0.05$ MeV (the uncertainty quoted by PDG) \cite{PDG}:
The uncertainty is $0.2 \%$. 
The branching ratio for $\phi \to K^+ K^-$ gives an 
uncertainty of $1.2\%$ following the PDG \cite{PDG}. 
The backgrounds from $N_{\phi K^{\pm} \nu}$ and 
$N_{\phi\pi^{\pm}\nu}$ have uncertainties of 0.3\% and 0.4\%,
respectively. The signal detection efficiency 
$\epsilon_{\phi K\nu}$ has an uncertainty of 0.5\%. 
A total systematic uncertainty of 6.5\% is obtained 
by adding all uncertainties in quadrature. 
The resulting branching fraction is then 
\begin{eqnarray}
B(\tau^{\pm} \to \phi K^{\pm}\nu) = (4.06 \pm 0.25 \pm 0.26) \times 10^{-5}. 
\end{eqnarray}
The systematic uncertainties for $\tau^{\pm}\to\phi K^{\pm}\pi^0\nu$ are
similar to those for $\tau^{\pm}\to\phi K^{\pm}\nu$. 
The main differences are on the trigger efficiency of 0.3\%, the detection efficiency 
$\epsilon_{\phi K\pi\nu}$ of 1.8\% and the $\pi^0$ detection efficiency of 1.7\%. 
Those provide a total systematic uncertainty of 6.9\%, and the branching 
fraction of 
\begin{eqnarray}
B(\tau^{\pm} \to \phi K^{\pm}\pi^0\nu) = (2.9 \pm 1.3 \pm 0.2) \times 10^{-6}. 
\end{eqnarray}

\begin{figure}[htb]
\centerline{\resizebox{10cm}{!}{\includegraphics{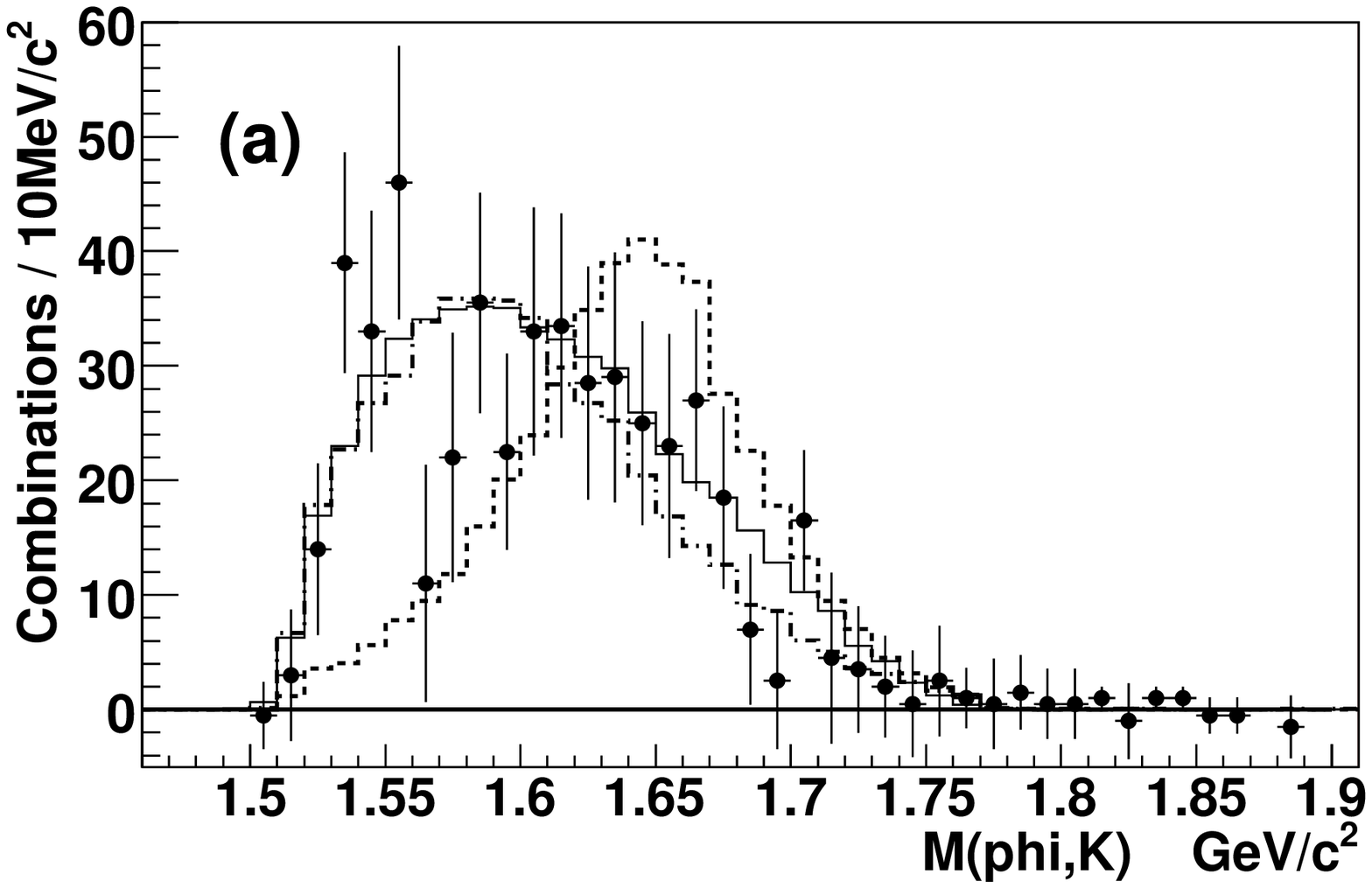}}}
\centerline{\resizebox{10cm}{!}{\includegraphics{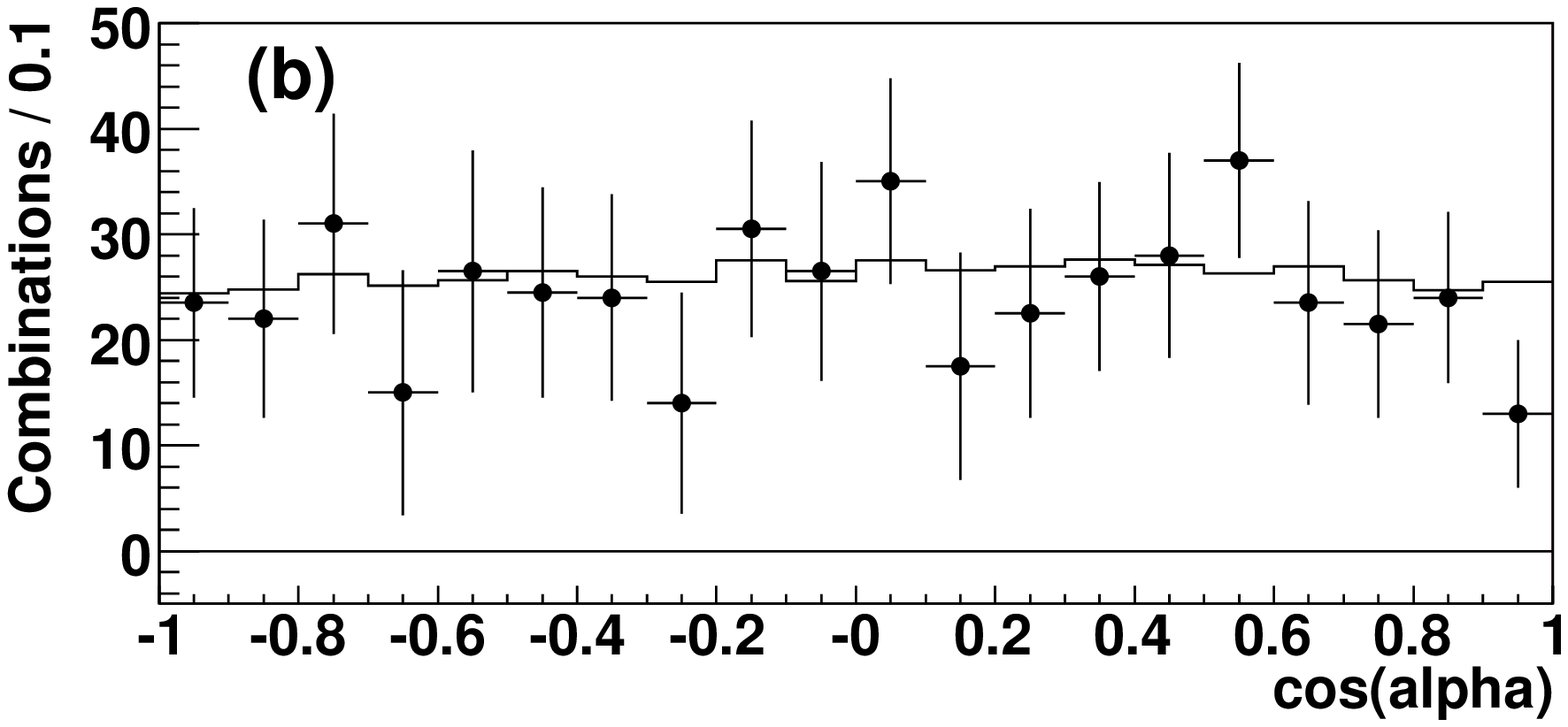}}}
\caption{(a) invariant mass and (b) angular distributions for 
$\phi K^{\pm}$ system. The non-$\phi$-resonant
backgrounds are subtracted using the sideband spectra.
Points with error bars indicate the data. 
Open histogram shows the phase space distributed signal MC, and 
dotted and dot-dashed histograms indicate the signal MC mediated by a resonance 
with $M=1650$ MeV/$c^2$ and $\Gamma=100$ MeV/$c^2$ and 
$M=1570$ MeV/$c^2$ and $\Gamma=150$ MeV/$c^2$, respectively.
In MC, the branching ratio of $4\times 10^{-5}$ is assumed.
(b) $\phi$'s angular distribution in the $\phi K^{\pm}$ rest frame,
where the direction of $P(\phi+K)$ in the laboratory frame 
is taken as the reference axis.
}
\label{fig:pm3}
\end{figure}

Finally, we examine a possible resonance state that intermediates the final $\phi K^{\pm}$ hadronic system. 
CLEO~\cite{CLEO} assumed a resonance having 
a mass of 1650 MeV/$c^2$ with a width 
of 100 MeV/$c^2$ in the evaluation of their detection 
efficiency for 
$\tau^{\pm}\to\phi K^{\pm}\nu$, however no signal was found. 
We generate a resonant MC with the KKMC simulation program. 
The weak current is generated with a $V-A$ form while
the $\phi K^\pm$ system is assumed to be produced from a 2-body decay of a
resonance.
In Fig.~\ref{fig:pm3}(a), the $\phi K^{\pm}$ mass distribution 
for data is compared to MC;
the combinatorial background is subtracted using the $K^+ K^-$ sideband.
Fig.~\ref{fig:pm3}(b) shows the $\phi$'s angular distribution 
in the $\phi K^{\pm}$ rest frame ($\cos \alpha$),
where the momentum direction of 
$\phi K^{\pm}$ in the laboratory frame ($P(\phi+K)$)
is taken as the reference axis. 
It indicates an isotropic distribution in the $\phi K^{\pm}$ system. 
For both the invariant mass and angular distributions of $\phi K^{\pm}$ system, 
the phase space MC reproduces the signal distribution well.
On the other hand, the 1650 MeV/$c^2$ state,
indicated by the dotted histogram in Fig.~\ref{fig:pm3}(a), 
clearly cannot account for the entire signal.
Assuming resonant production, the best agreement with the data
is found for a mass and a width of
$\simeq$1570 MeV/$c^2$ and $\simeq$150 MeV/$c^2$, respectively, as shown 
by the dot-dashed histogram. 
However, since the shape of the resonant MC is similar to
the phase space distributed MC,
it is inappropriate to look for the 
intermediate resonance state with $\Gamma \sim O(100{\rm MeV})$
in this narrow mass range of $\sim$250 MeV/$c^2$.
In fact, the phase space distribution (the open histogram)
agrees well with data. 

\section{Conclusion}

Using a 401 fb$^{-1}$ data sample, we report the first 
observation of the rare $\tau$ decay mode,
$\tau^{\pm} \to \phi K^{\pm}\nu$, with a branching fraction of 
\begin{eqnarray}
B(\tau^{\pm} \to \phi K^{\pm}\nu) = (4.06 \pm 0.25 \pm 0.26) \times 10^{-5}. 
\end{eqnarray}

\smallskip
\bigskip
\noindent
{\bf Acknowledgments}
\smallskip

We would like to gratefully acknowledge the essential
contributions of Mari Kitayabu, which are described in her bachelor thesis 
at Nagoya University.
We thank the KEKB group for the excellent operation of the
accelerator, the KEK cryogenics group for the efficient
operation of the solenoid, and the KEK computer group and
the National Institute of Informatics for valuable computing
and Super-SINET network support. We acknowledge support from
the Ministry of Education, Culture, Sports, Science, and
Technology of Japan and the Japan Society for the Promotion
of Science; the Australian Research Council and the
Australian Department of Education, Science and Training;
the National Science Foundation of China and the Knowledge
Innovation Program of the Chinese Academy of Sciencies under
contract No.~10575109 and IHEP-U-503; the Department of
Science and Technology of India; 
the BK21 program of the Ministry of Education of Korea, 
the CHEP SRC program and Basic Research program 
(grant No.~R01-2005-000-10089-0) of the Korea Science and
Engineering Foundation, and the Pure Basic Research Group 
program of the Korea Research Foundation; 
the Polish State Committee for Scientific Research; 
the Ministry of Science and Technology of the Russian
Federation; the Slovenian Research Agency;  the Swiss
National Science Foundation; the National Science Council
and the Ministry of Education of Taiwan; and the U.S.\
Department of Energy.

\end{document}